\documentstyle[times,afd-gen,afd-orm,asymetrix,url,sq]{article}

\title{Interactive Query Formulation using 
       Spider Queries\\
       {\normalsize Asymetrix Report 94-2}
}
\author{
   H.A. Proper\\
   Asymetrix Research Laboratory\\
   Department of Computer Science\\
   University of Queensland\\
   Australia 4072\\
   E.Proper@acm.org
}
\date{\Version}

   \input{epsf}
   \def\Figurepath{./}
   \def\Scale{0.9}
   \def\epsfsize#./##2{\Scale#./}

\begin{document}
   \maketitle
   {\sc Published as:}
\begin{quote}
  H.A.~(Erik) {Proper}. {Interactive Query Formulation using Spider Queries}. Technical report, Asymetrix Research Laboratory, University of Queensland, Brisbane, Queensland, Australia, 1994.
\end{quote}

   \begin{abstract}
   Effective information disclosure in the context of databases with a large conceptual 
   schema is known to be a non-trivial problem. In particular the formulation of ad-hoc 
   queries is a major problem in such contexts. Existing approaches for tackling this 
   problem include graphical query interfaces, query by navigation, query by construction, 
   and point to point queries. In this article we propose the spider query mechanism as a 
   final corner stone for an easy to use computer supported query formulation mechanism for 
   InfoAssisant.

   The basic idea behind a spider query is to build a (partial) query of all information
   considered to be relevant with respect to a given object type. 
   The result of this process is always a tree that fans out over existing conceptual 
   schema (a spider).
  
   We also provide a brief discussion on the integration of the spider quer mechanism with 
   the existing query by navigation, query by construction, and point to point query 
   mechanisms.
\end{abstract}

   \section{Introduction}
Most present day organisations make use of some automated information system. This usually means 
that a large body of vital corporate information is stored in these information systems. As a result an 
essential function of information systems is the support of disclosure of this 
information. Without a set of adequate information disclosure avenues an information system becomes 
worthless since there is no use in storing information that will never be retrieved. An adequate support 
for information disclosure, however, is far from a trivial problem. Most query languages do not 
provide any support for the users in their quest for information. Furthermore, the conceptual schemata 
of real-life applications tend to be quite large and complicated. As a result, the users may easily 
become 'lost in conceptual space' and they will end up retrieving irrelevant (or even wrong) objects 
and may miss out on relevant objects. Retrieving irrelevant objects leads to a low precision, missing 
relevant objects has a negative impact on the {\em recall} (\cite{Book:83:Salton:IntroIR}).

The disclosure of information stored in an information system has some clear parallels to the 
disclosure problems encountered in {\em document retrieval systems}. To draw this parallel in more 
detail, we quote the information retrieval paradigm as introduced in 
\cite{Report:91:Bruza:StratHypmed}. The paradigm starts with an individual or company having an 
{\em information need} they wish to fulfil. This need is typically a vague notion and needs to be made 
more concrete in terms of an {\em information request} (the query) in some (formal) language. The 
information request should be as good as possible a description of the information need. The 
information request is then passed on to an automated system, or a human intermediary, who will 
then try to fulfil the information request using the information stored in the system. This is illustrated 
in the {\em information disclosure}, or {\em information retrieval paradigm}, presented in 
\SRef{\IRParadigm} which is taken from \cite{Report:91:Bruza:StratHypmed}.

We now briefly discuss why the information retrieval paradigm for document retrieval systems is also 
applicable for information systems. For a more elaborate discussion on the relation between 
information systems and document (information) retrieval systems in the context of the information 
retrieval paradigm, refer to \cite{PhdThesis:94:Proper:EvolvConcModels}. In the paradigm, the 
retrievable information is modelled as a set \Cal{K} of {\em information objects} constituting the 
{\em information base} (or population).

In a document retrieval system the information base will be a set of documents 
(\cite{Book:83:Salton:IntroIR}), while in the case of an information system the information base will 
contain a set of facts conforming to a conceptual schema. Each information object $o \in \Cal{K}$ is 
{\em characterised} by a set of descriptors $\Cal{X}(o)$ that facilitates its disclosure. The 
characterisation of information objects is carried out by a process referred to as indexing. In an 
information system, the stored objects (the population or information base) can always be identified by 
a set of (denotable) values, the identification of the object. For example, an address may be identified 
as a city name, street name, and house number. The characterisation of objects in an information 
system is directly provided by the reference schemes of the object types. 

The actual information disclosure is driven by a process referred to as {\em matching}. In document 
retrieval applications this matching process tends to be rather complex. The characterisation of 
documents is known to be a hard problem (\cite{Article:77:Maron:InfRetr}, 
\cite{Book:86:Craven:SringIndexing}), although newly developed approaches turn out to be quite 
successful (\cite{Book:89:Salton:AutomTextProc}). In information systems the matching process is 
less complex as the objects in the information base have a more clear characterisation (the 
identification). In this case, the identification of the objects (facts) is simply related to the query 
formulation $q$ by some (formal) query language.
{\def\Scale{0.5} \EpsfFig[\IRParadigm]{The information retrieval paradigm}}

The remaining problem is the query formulation process itself. An easy and intuitive way to formulate 
queries is absolutely essential for an adequate information disclosure. Quite often, the quest from users 
to fulfil their information need can be aptly described by (\cite{PhdThesis:92:Bruza:IRHypmed}):
\begin{quote} \it 
   I don't know what I'm looking for, but I'll know when I find it.
\end{quote}
In document retrieval systems this problem is attacked by using {\em query by navigation} 
(\cite{Report:91:Bruza:StratHypmed}, \cite{PhdThesis:92:Bruza:IRHypmed}) and {\em relevance 
feedback} mechanisms (\cite{Article:89:Rijsbergen:IRLogic}). The query by navigation interaction 
mechanism between a searcher and the system is well-known from the Information Retrieval field, 
and has proven to be useful. It shall come as no surprise that these mechanisms also apply to the query 
formulation problem for information systems. In \cite{Report:92:Burgers:PSMIR}, 
\cite{Report:93:Burgers:PSMIR}, \cite{Report:93:Hofstede:DisclSupport}, 
\cite{PhdThesis:94:Proper:EvolvConcModels} such applications of the {\em query by navigation} 
and {\em relevance feedback} mechanisms have been described before. When combining the query by 
navigation and manipulation mechanisms with the ideas behind visual interfaces for query 
formulation as described in e.g. \cite{Article:92:Auddino:SUPERVisual} and 
\cite{Article:94:Rosengren:VisualER} powerfull and intuitive tools for computer supported query 
formulation become feasible. Such tools will also heavily rely on the ideas of direct manipulation 
interfaces (\cite{Article:83:Shneiderman:DirectManip}) as used in present day computer interfaces. 

One important step in the improvement of the information disclosure of information systems, is the 
introduction of query languages on a conceptual level. Examples of such conceptual query languages 
are RIDL (\cite{Report:82:Meersman:RIDL}), LISA-D (\cite{Report:91:Hofstede:LISA-D}, 
\cite{Report:92:Hofstede:LISA-DPromo}), and FORML (\cite{Article:92:Halpin:Subtyping}). By 
letting users formulate queries on a conceptual level, users are safeguarded from having to know the 
exact mapping to internal representations (e.g. a set of tables which conform to the 
relational model) to be able to formulate queries in a non conceptual language such as SQL. 
The next step is to introduce ways to support users in the formulation of queries in such conceptual 
query languages (CQL).

In line with the above discussed information retrieval paradigm and the notion of relevance feedback, 
a query formulation process (both for a document retrieval system, and an information system) can be 
said to roughly consist of the following four phases:
\begin{enumerate}
   \item The {\em explorative phase}. What information is there, and what does it mean?
   \item The {\em constructive phase}. Using the results of phase 1, the actual query is formulated.
   \item The {\em feedback phase}. The result from the query formulated in phase 2 may not be completely
         satisfactory. In this case, phases 1 and 2 need to be re-done and the result refined.
   \item The {\em presentation phase}. In most cases, the result of a query needs to be incorporated into a
         report or some other document. This means that the results must be grouped or aggregated in some
         form. 
\end{enumerate}
Depending on the user's knowledge of the system, the importance of the respective phases may 
change. For instance, a user who has a good working knowledge of the structure of the stored 
information may not require an elaborate first phase and would like to proceed with the second phase 
as soon as possible. 

In this report, we discuss an additional mechanism to support automated disclosure of 
information stored in information systems, the spider query mechanism. 
As stated before, the related notions of {\em query by navigation} and 
{\em query by construction} have already been discussed in 
\cite{Report:92:Burgers:PSMIR}, \cite{Report:93:Proper:DisclSch}, 
\cite{PhdThesis:94:Proper:EvolvConcModels}. 
The point to point query mechanism was already discussed in \cite{AsyReport:94:Proper:PPQ}. 

The idea behind spider queries is to start out from one object type, and to associate all 
information that is relevant to this object type. 
The essential part of a spider query is selecting the object types in the direct suroundings 
of the initial object type that are considered to be relevant. 
This style of querying corresponds to a situation where users only know about the existance of 
some object types in the conceptual schema about which they would like to be informed. 

The structure of this report is as follows. 
In \SRef{SampleSession}, we discuss an example spider query session, and elaborate briefly on the
integration with the existing query by navigation, query by construction, and point to point querie
mechanisms. Section \ref{GraphRepr} deals with the representation of conceptual schema as a graph. 
Building a spider query (essentailly also a graph) is covered in \SRef{SpiderQuery}. Before 
concluding, section \SRef{SpiderPath} discusses the representation of a spider query as a 
path expression.
For the reader who is unfamiliar with the notation style used in this report,
it is advisable to first read \cite{AsyReport:94:Proper:Formal}.

   \section{An Example Spider Query Session}
\SLabel{section}{SampleSession}

In this section we discuss a sample session involving a spider query, and also discuss briefly the 
relationship to the existing query by navigation, query by construction and point to point queries. The 
discussed example operates on a conceptual schema for the administration of the election of American 
presidents. The example schema itself is not shown; the structure of the domain will become clear 
from the sample session. Note that the quality of the verbalisations of the paths in the examples used 
in this section should be improved, however, this is subject of further research. In \SRef{\SQStart} a 
possible screen is depicted for building queries using a point to point query mechanism. No special 
window is needed for a spider query (see also \cite{AsyReport:94:Proper:PPQ}).
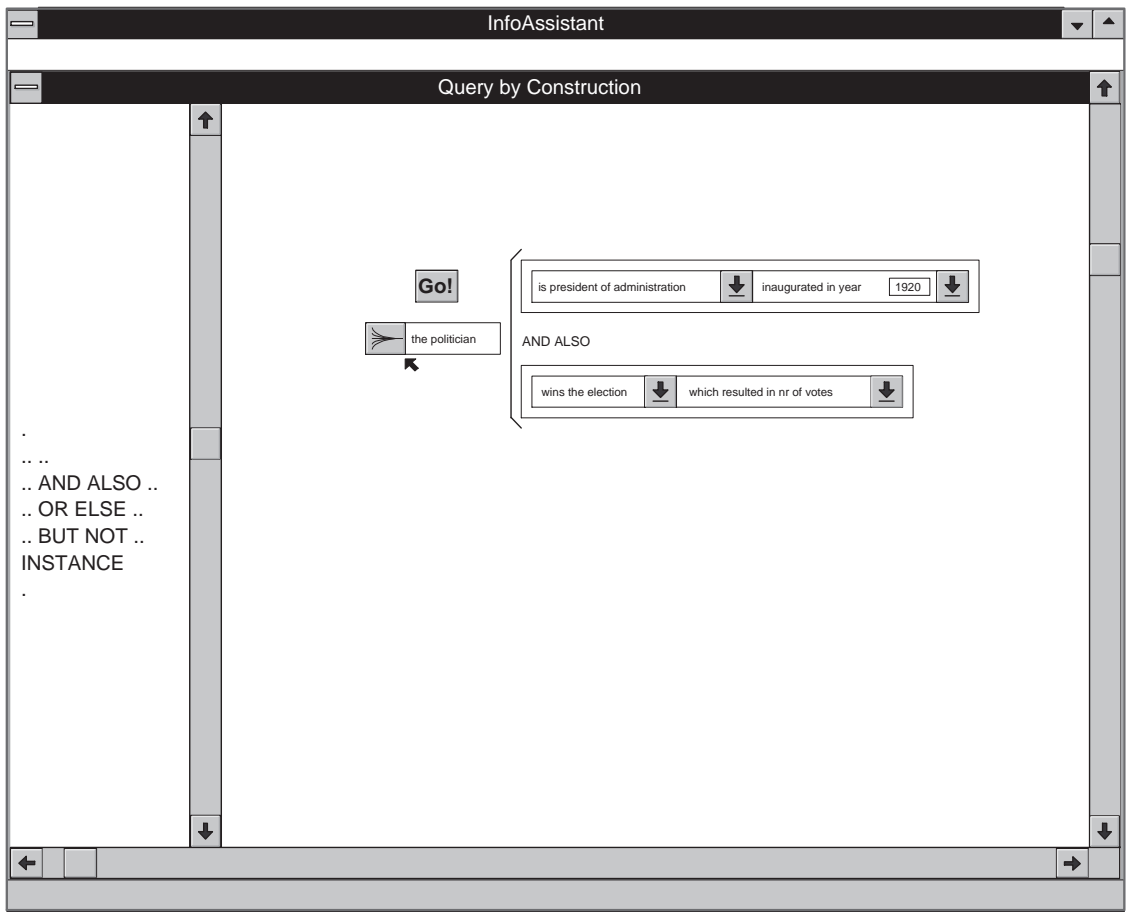
{\def\Scale{0.70} \EpsfFig[\SQStart]{Start of a spider query}}

We start out from an existing query in a query by construction window. Note that this could also be 
single object type, e.g. politician. The spider query mechanism adds one important aspect to the query 
by construction window, the spider button: \SpiderButton. When a user presses this button, the system 
calculates the spider query of the object type directly to the right of the button. This is illustrated in 
\SRef{\SQResult}. 
The system allows for the removal of parts of the resulting spider query that are not considered 
to be relevant by the user.
Suppose the user is not interested in \SF{administration is headed by} and 
\SF{election won by}, then these paths can be deleted, which leads to the screen depicted in
\SRef{\SQResultPruned}.
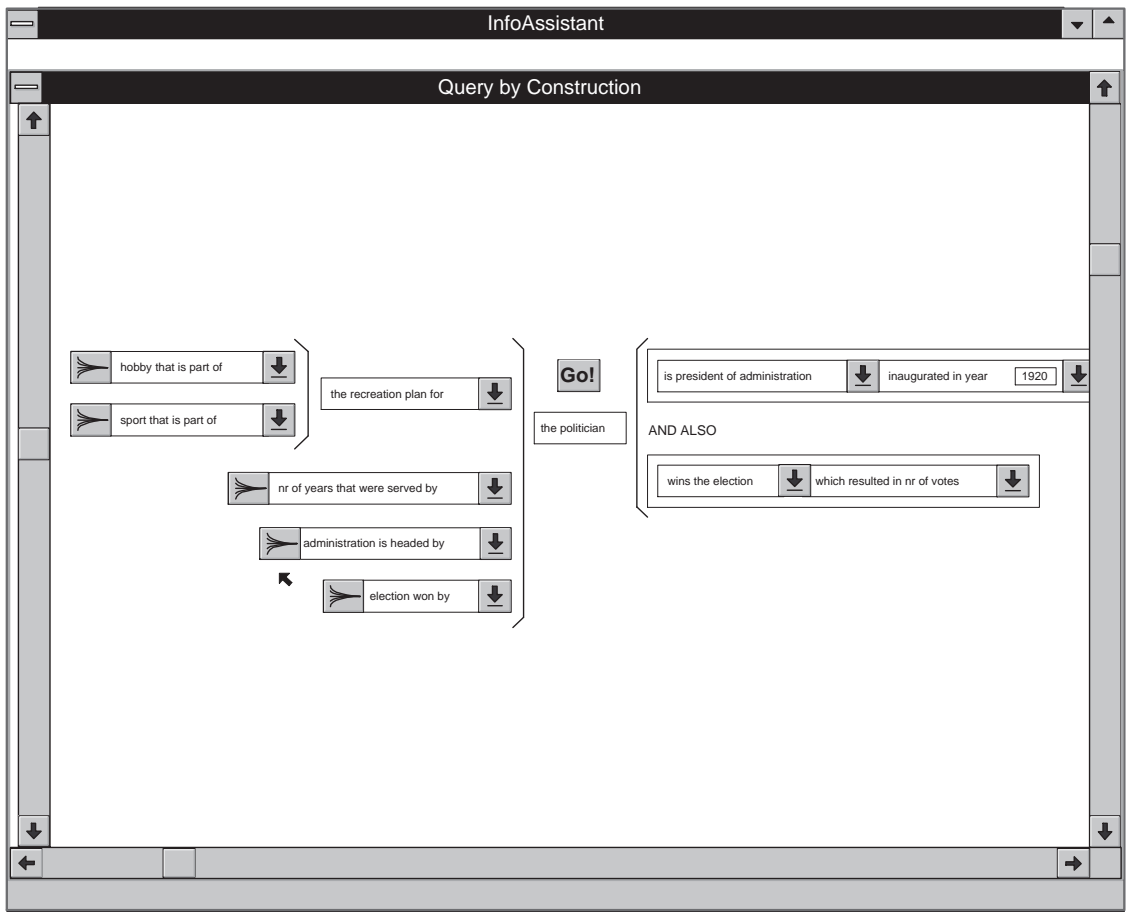
{\def\Scale{0.70} \EpsfFig[\SQResult]{Result of a Spider Query}}

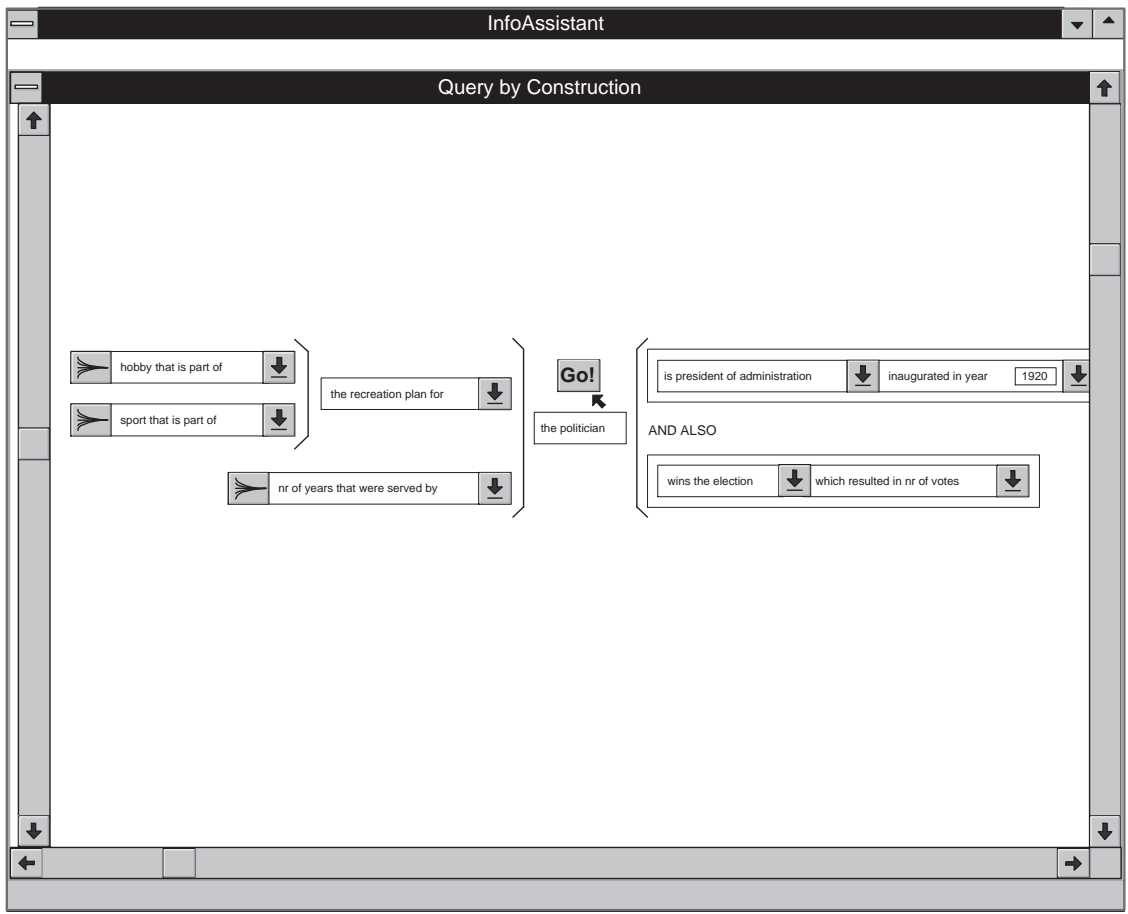
{\def\Scale{0.70} \EpsfFig[\SQResultPruned]{Pruning the Spider Query}}
It is now interesting to see that a query essentially is a double tree with a shared root
(politician in the example).
Furthermore, the leaves on the tree resulting from the spider query can be extended further 
if desired by commencing new spider queries.
Finally, since the result of a spider query is constructed from path expressions as well, these 
expressions have the \PPQButton~associated that can be used to select alternative paths between the 
head and tail object types. Furthermore, the paths can also be used as a starting point of a query by 
navigation session. 
This latter posibility is illustrated in \SRef{\SQQBN}.
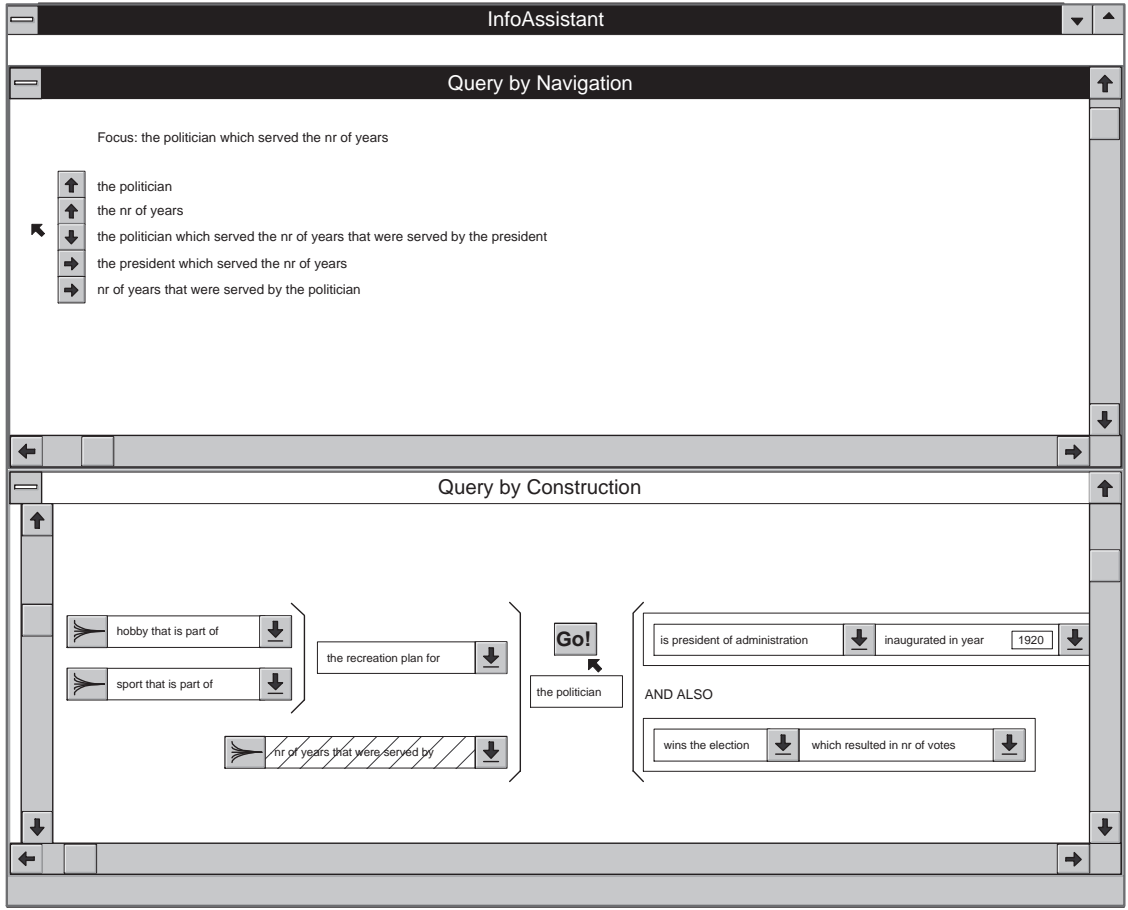
{\def\Scale{0.70} \EpsfFig[\SQQBN]{Switching to query by navigation}}

As stated before in \cite{AsyReport:94:Proper:PPQ}, the query by construction window is basically a syntax directed editor. In 
the left part of the window all possible constructs from the query language are listed. In our examples 
we have used the constructs defined in LISA-D. Once the FORML and LISA-D languages have been 
merged, a more complete language for the query by construction part will result.

   \section{A Conceptual Schema as a Graph}
\SLabel{section}{GraphRepr}
For the purpose of finding a path between object types in a conceptual schema, 
the schema first needs to be translated to a graph. 
This translation is exactly the same as provided in \cite{AsyReport:94:Proper:PPQ}, but for 
reasons of completeness we provide it again. 
We start out from a formalisation of ORM based on the one 
used in (\cite{Report:94:Halpin:ORMPoly}). 
However, since only a very limited part of the 
formalisation is needed, we do not cover the formalisation in full detail.

A conceptual schema is presumed to consist of a set of types $\Types$. 
Within this set of types two subsets can be distinguished: the relationship types 
$\RelTypes$, and the object types $\ObjTypes$. 
Furthermore, let $\Preds$ be the set of roles in the conceptual schema. 
The fabric of the conceptual schema is then captured by two functions and two predicates. 
The set of roles associated to a relationship type are provided by the partition: 
$\Roles: \RelTypes \Func \Powerset(\Preds)$. 
Using this partition, we can define the function $\Rel$ which returns for each role
the relationship type in which it is involved:
   $\Rel(r) = f \iff r \in \Roles(f)$.
Every role has an object type at its base called the player of the role, which is 
provided by the function: $\Player: \Preds \Func \Types$. 
Subtyping and polymorphy of object types are captured by the predicates 
$\Spec \subseteq \ObjTypes \Carth \ObjTypes$ and 
$\Poly \subseteq \ObjTypes \Carth \ObjTypes$ 
respectively. 
For any ORM conceptual schema the following (undirected) labelled graph 
$G = \tuple{N,E}$ can then be defined:
\begin{eqnarray}
   N & \Eq    & \Types\\
   E & \Eq    & \Set{\tuple{\setje{\Player(r),\Rel(r)}, r}}{r \in \Preds}\\
     & \Union & \Set{\tuple{\setje{x,y},\Spec}}{x \Spec y}\\
     & \Union & \Set{\tuple{\setje{x,y},\Poly}}{x \Poly y}
\end{eqnarray}
The edges in the resulting graph have the format $\tuple{\setje{x,y},l}$, where $x$ and $y$ are the 
source/destination (no order) of the edge, and $l$ is the label of the edge. 
The labels on the edges either result from the roles in the relationship types (2), 
or they result from specialisation or polymorphism (3,4).
In the remainder, the graph 
$G$ will be used as an implicit parameter for all introduced functions and operations. As a 
convention, the nodes of graph $G$ are accessed by $G.N$, and the edges by $G.E$.
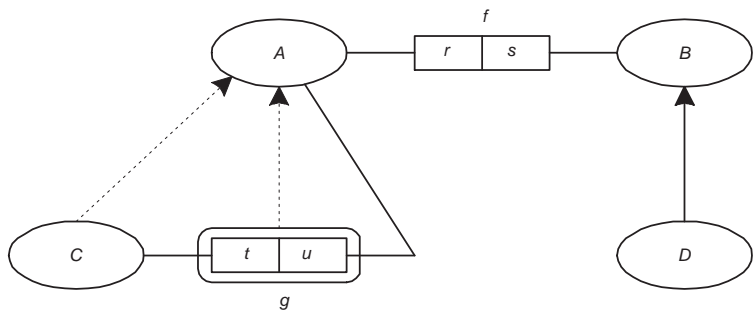
\EpsfFig[\ExCS]{Example Conceptual Schema}

As an example, consider the conceptual schema depicted in \SRef{\ExCS}. For this schema we have:
\[ \begin{array}{ll}
   \Types    = \setje{A,B,C,D,f,g} & \Roles(f)=\setje{r,s}, \Roles(g)=\setje{t,u}\Eol
   \RelTypes = \setje{f,g}         & \Player(r)=A, \Player(s)=B, \Player(t)=C, \Player(u)=A\Eol
   \ObjTypes = \setje{A,B,C,D}     & A \Poly C, A \Poly g\Eol
   \Preds    = \setje{r,s,t,u}     & D \Spec B
\end{array} \]
From this schema the graph as depicted in \SRef{\ExGraph} can be derived.
\EpsfFig[\ExGraph]{Example Graph}

Note that for spider queries, it might be usefull to limit the edges resulting 
from roles to those roles which are {\em not} used in the reference 
schemas of other object types. 
We cannot yet decide on this until we have finallised the path expression language, but it
is simply a matter of filtering the proper edges.

   \section{Building a Spider Query}
\SLabel{section}{SpiderQuery}

The construction of a spider query corresponds to the construction of 
a graph.
The nodes and edges in this graph are based on the nodes (object types) and
edges from the original conceptual schema graph.
A {\em spider query graph} is a labelled tree (connected directed acyclic 
graph).
It is not just a subgraph of the conceptual schema graph, since one 
object type can be visited more than once on different branches of the graph 
(legs of the spider query graph). 
Let $\Nodes$ be the set of all nodes that can occur in a 
spider query, and let $\Labels \Eq \Preds \union \setje{\Spec,\Poly}$ be the set of
labels that can occur in a spider query graph.
The spider query graph itself is now constructed by the function:
\[ \SpiderQuery: \Nodes ~\Func~ 
   (\Nodes \PartFunc \Types) \Carth 
   \Powerset(\Nodes \Carth \Nodes \Carth \Labels) 
\]
The result of a spider query $\SpiderQuery(x)$ is a tuple 
$\tuple{O,S}$, where $O: \Nodes \PartFunc \Types$ provides the relation between 
the spider query graph and the conceptual schema, and the spider query graph itself
is defined by the (directed!) edges in 
$S \subseteq \Powerset(\Nodes \Carth \Nodes \Carth \Labels)$.

The construction process itself is driven by a recursive function $\sigma$, 
which is activated as follows:
\[ \SpiderQuery(x) ~\Eq~ \sigma(\setje{\tuple{\SF{NewNode}(x),x}},\emptyset,\setje{x}) \]
where $\SF{NewNode}$ is a function returning a new node each time it is called.
The $\sigma$ function is the actual engine of the construction process.
This function tries to extend the spider query graph in a number of 
steps.
In each step the possible extensions of the existing graph at that moment are 
calculated by the $\varepsilon$ function.
The $\varepsilon$ function takes as parameters the current spider query 
graph: $O$, $S$, and the nodes which are allowed to be extended: $T$.
It returns a number of tuples of the form $\tuple{n,t,l}$, where $n$ is an 
extendable node in the existing spider query graph, and $t$ is a type in the 
conceptual schema that is reachable in (the conceptual schema graph) from 
object type $O(n)$.
A restriction on the returned tuples is that no cycles may be formed, i.e. no revisiting of 
object types on one path in the spider query graph. 
The formal definition is given by:
\[ \varepsilon(O,S,T) ~\Eq~ 
   \Set{\tuple{n,t,l}}{
      \tuple{\setje{O(n),t},l} \in G'.E \land n \in T \land t \not\in \Top(n)
   }
\]    
where $\Top(n) \Eq \Union\Sub{m: \tuple{m,n,l} \in S} (\Top(m) \union \setje{m})$ is the
set of types on the path in the spider query graph leading from the root to $n$.

The actual driver function $\sigma$ evolves around three parameters.
These parameters are updated in every step of the function.
In the definition we use $O$, $S$ and $T$ as variable names, where $T$ contains the 
nodes in the spider query graph that may be used for further extensions, and $O$ and $S$ 
represent the spider query graph so-far.
The $\sigma$ function is identified by:
\[ \sigma(O,S,T) ~\Eq~
   \left\{ \begin{array}{ll}
      \sigma(O',S',T') & \mbox{if~} \varepsilon(O,S,T) \neq \emptyset\\
      \tuple{O,S}      & \mbox{otherwise}
   \end{array} \right. 
\]
where:
\begin{eqnarray}
   O' &=& O \union \Set{\tuple{\SF{NewNode}(t),t}}{t \in \pi_2~\varepsilon(O,S,T)}\\
   S' &=& S \union \Set{\tuple{n,m,l}}{\tuple{n,t,l} \in \varepsilon(O,S,T) \land O'(m)=t}\\
   T' &=& \Set{m}{\tuple{n,m,l} \in \varepsilon(O,S,T) \land \CWeight(O'(m)) \leq \CWeight(O'(n))}
\end{eqnarray}
In 4, the newly found object types in $\varepsilon(O,S,T)$ are assigned a new node so that
they can be added to the existing spider query graph.
The set of edges of the spider query graph is updated in 5.
The new set of nodes that will be considered for further extensions in the next step of 
$\sigma$ are determined by 6.
In this definition, the conceptual weight function $\CWeight$ is the same function as used 
in \cite{AsyReport:94:Proper:PPQ}, and should provide the conceptual importance of each object type. 
This importance could for instance be based on the abstraction level at which 
the object type occurs (\cite{Article:94:Campbell:Abstraction}).
The rationale behind the use of the $\CWeight$ function is that as soon as the conceptual 
importance increases, any new neigbouring object type from the last added object type 
(the one with the increased $\CWeight$) is not relevant for the root of the current spider 
query graph, i.e. we have left the relevance scope of the current root. 
Note, however, the neighbours of the last added object type could quite well be relevant for
a spider query starting out from this latter object type. 
These nodes will be added if the user presses the spider query button that will be associated
to this node.

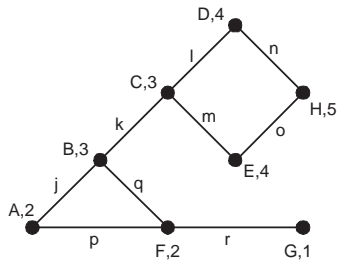
\EpsfFig[\ExSQGraph]{Example graph for a spider query}
As an example of the operation of the $\SpiderQuery$ function, consider the graph depicted
in \SRef{\ExSQGraph}.
Each edge in this graph is labelled, and each node has associated its name (object type), and
the conceptual weight.
In \SRef{\ExSQA} the tree is depicted with which the $\sigma$ function is started.
The double circle around node \SF{B} is used to indicate that node \SF{B} is in the set
of extendible nodes $T$.
\EpsfFig[\ExSQA]{Initial spider query graph}

Figure \ref{\ExSQB} depicts the spider query graph after one incremental step of $\sigma$.
All neighbours of \SF{B} are added to the graph, and since they do not have a higher
conceptual weight than \SF{B}, they can be used for further extensions.
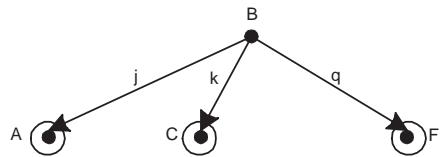
\EpsfFig[\ExSQB]{First step in building the spider query graph}

The next step of the algorithm is illustrated in \SRef{\ExSQC}.
Node \SF{A} has two neigbours: \SF{B} and \SF{F}.
However, since adding \SF{B} to the spider query graph would lead to a repetition of
an existing node on the path to the root of the spider query graph, \SF{B} is not added.
Similarly, \SF{B} is not added as a neighbour of \SF{C} and \SF{F}.
Nodes \SF{D} and \SF{E} are not marked as points of further extensions since they both have
a higher conceptual weight then node \SF{C}. 
As a resuolt, node \SF{H} is not part of node \SF{B}'s scope of rellevance.
However, node \SF{B} would be part of a spider query starting from \SF{H}.
Note that nodes \SF{D} and \SF{E} will both have associated a spider query button when
the result is presented to the user, so a user can always explicitly decide to further 
`climb the conceptual importance mountain'.
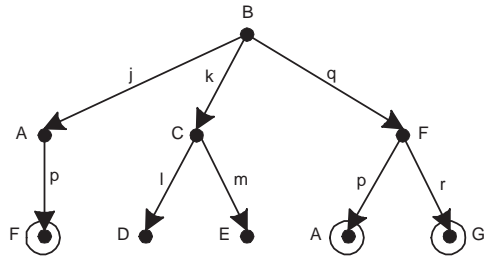
\EpsfFig[\ExSQC]{Second step in building the spider query graph}

The result of the last step of $\sigma$ is shown in \SRef{\ExSQD}.
The second node \SF{A}, and node \SF{G}, do not lead to further extensions. 
All neighbours of these two nodes are already present on the paths to the root
of the graph.
Only the leftmost node \SF{F} leads to a further extension with a \SF{G}.
After this extension no further extensions are possible. 
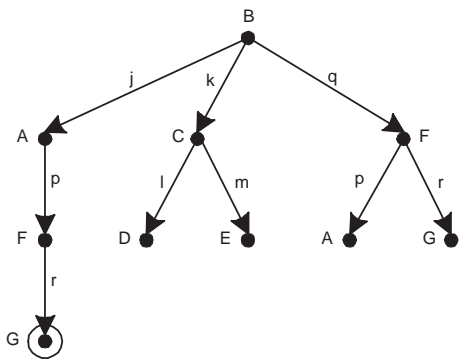
\EpsfFig[\ExSQD]{Last step in building the spider query graph}

Finally, it is good to realise that the $\sigma$ function always terminates:
\begin{lemma}
   The $\sigma$ function always terminates.
\end{lemma}
\begin{proof}
   This corresponds to saying that the resulting graph is finite.

   The number of object types in an ORM schema is (presumed to be) finite, and the paths of the spider 
   query will not contain cycles (follows from the definition of $\varepsilon$).

   As a result, each node only has a finite number of outgoing arcs, and each path from the 
   root to a leaf is finite. Hence the resulting graph is finite.
\end{proof}

   \section{The Resulting Path Expressions}
\SLabel{section}{SpiderPath}

In this section we discuss how to transfer a spider query graph into a 
path expression.
We use the spider query graph $O$, $S$ as an implicit parameter for all definitions in
this section.

Given a node $x$ in the spider query graph, then the following path expression can be 
associated to this node:
\[
  \SF{NodeExpr}(x) ~\Eq~
  \left\{ \begin{array}{ll}
      [a_1:\SF{PathSeg}(y_1,l_1,x), \ldots, 
       a_n:\SF{PathSeg}(y_n,l_n,x); O(x)] &
         \mbox{if~} S \neq \emptyset\Eol
      O(x) & \mbox{otherwise}
  \end{array} \right.
\]
where $S = \setje{\tuple{y_1,l_1}, \ldots, \tuple{y_n,l_n}}$ is a set such 
that $S = \Set{\tuple{y,l}}{\tuple{x,y,l} \in S}$, and $a_1,\ldots,a_n$
is a set of fresh attribute names.
A good choice for these latter names are the names of the object types where the
$\SF{PathSeg}$s end (suffixed with a number to make the name unique if needed).
The $\Conc$ operation is the concatenation operation for path expressions, 
and the $[X_1, \ldots, X_n]$ construct is the path confluence operation. 
It allows us to combine a variety of path expressions.
For more details about the path expression operators, refer to
\cite{Report:91:Hofstede:LISA-D} and the forthcomming Asymetrix report on 
path expressions.
One single edge from the spider query graph is converted to a path expression 
as follows:
\[ 
  \SF{PathSeg}(y,l,x) ~\Eq~ \SF{NodeExpr}(y)~\SF{Connector}(l,x)~O(x)
\] 
where
\[ 
   \SF{Connector}(l,x) ~\Eq~
      \left\{ \begin{array}{ll}
        \Conc               & \mbox{if~} l \in \setje{\Poly,\Spec}\Eol
        \Conc l       \Conc & \mbox{if~} x \in \RelTypes \land l \in \Roles(x)\Eol
        \Conc \Rev{l} \Conc & \mbox{otherwise}
   \end{array} \right.
\] 

The linear path expressions are for internal use only. 
They can be mapped to proper SQL queries on the one hand, and verbalised as 
semi-natural language sentences using the verbalisation information as 
provided in the conceptual schema on the other hand. 
As stated before, the verbalisation of path expressions is subject of further 
research.

Finally, the (unique) root of a spider query graph is determined as follows:
\[ \SF{IsRoot}(r) \iff \DefinedAt{O}{r} \land \lnot\Ex{x,l}{\tuple{x,r,l} \in S} \]
If $r$ is the (unique) root of a spider query graph, then $\SF{NodeExpr}(r)$ results in
the complete path expression for this spider query graph.

   \section{Conclusions}

In this article we introduced a novel way to computer supported query 
formulation called spider queries. 
We provided a sample session with a provisional tool supporting spider queries,
and briefly discussed the relationship to query by navigation, query by 
construction, and point to point queries. 
Together with these existing mechanisms a powerfull query formulation tool 
can now indeed be build. 

As a next step, the path expressions should be further developed to suit our 
needs. 
Furthermore, elegant verbalisations of the path expressions should be catered 
for.

   \AddBib{asy}
   \BIBLIOGRAPHY{alpha}
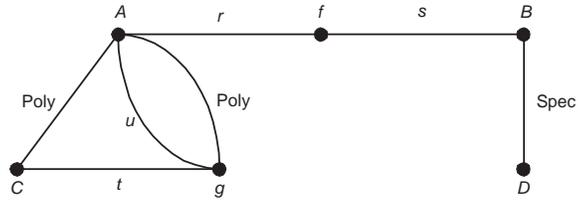
\end{document}